\documentstyle[11pt]{article}
\begin{document}
\input epsf
\input psbox.tex
\psfordvips

%%%%%%%%%%%%%%%%%%%%%%%%%%%%%%%%%%%%%%%%%%%%%%%%%%%%%%%%%%%%%%%%%%%%%%%%
%%%%%%%%%%%%%%%%%%%%%%%%%%%%%%%%%%%%%%%%%%%%%%%%%%%%%%%%%%%%%%%%%%%%%%%%

\def\com#1,#2,#3#4{          {\it Comm. Math. Phys.\/ }{\bf #1} (19#3#4) #2}
\def\np#1,#2,#3#4{           {\it Nucl. Phys.\/ }{\bf B#1} (19#3#4) #2}
\def\npps#1,#2,#3#4{         {\it Nucl. Phys. B (Proc. Suppl.)\/ }{\bf B#1}
                             (19#3#4) #2}
\def\pl#1,#2,#3#4{           {\it Phys. Lett.\/ }{\bf #1B} (19#3#4) #2}
\def\pla#1,#2,#3#4{          {\it Phys. Lett.\/ }{\bf #1A} (19#3#4) #2}
\def\pr#1,#2,#3#4{           {\it Phys. Rev.\/ }{\bf D#1} (19#3#4) #2}
\def\prep#1,#2,#3#4{         {\it Phys. Rep.\/ }{\bf #1} (19#3#4) #2}
\def\prl#1,#2,#3#4{          {\it Phys. Rev. Lett.\/ }{\bf #1} (19#3#4) #2}
\def\pro#1,#2,#3#4{          {\it Prog. Theor. Phys.\/ }{\bf #1} (19#3#4) #2}
\def\rmp#1,#2,#3#4{          {\it Rev. Mod. Phys.\/ }{\bf #1} (19#3#4) #2}
\def\sp#1,#2,#3#4{           {\it Sov. Phys.-Usp.\/ }{\bf #1} (19#3#4) #2}
%%%%%%%%%%%%%%%%%%%%%%%%%%%%%%%%%%%%%%%%%%%%%%%%%%%%%%%%%%%%%%%%%%%%%%%%%%%%

\def\thefootnote{\fnsymbol{footnote}}
\def\integer{\rm integer}
\def\R#1{R^{#1}}
\def\pr#1{\phi_{#1}}
\def\mod{\rm mod}
\def\s#1#2{{\tilde S_{#1,#2}}}
\def\draft{{\hskip 4cm {\Large DRAFT}}}

%%%%%%%%%%%%%%%%%%%%%%%%%%%%%%%%%%%%%%%%%%%%%%%%%%%%%%%%%%%%%%%%%%%%%%%%%%%%
\begin{center}
\hfill    WIS-96/6/Jan-PH\\
\hfill       hep-th/9601114
\vskip 1 cm

{\large \bf  Fusion Rules for Extended Current Algebras}

\vskip 1 cm

Ernest Baver and Doron Gepner

\vskip 1 cm

{\em Department of Particle Physics\\
The Weizmann Institute\\
Rehovot 76100\\
ISRAEL}

\end{center}

\vskip 1 cm

\begin{abstract}

The initial classification of fusion rules have shown that rational conformal field theory is very limited. In this paper we study the fusion rules of extended current algebras. Explicit formulas are given for the S matrix and the fusion rules, based on the full splitting of the fixed point fields. We find that in some cases sensible fusion rules are obtained, while in others this procedure leads to fractional fusion constants.

\end{abstract}

\newpage

%%%%%%%%%%%%%%%%%%%%%%%%%%%%%%%%%%%%%%%%%%%%%%%%%%%%%%%%%%%%%%%%%%%%%%%%%%%%

%\draft
\vskip 1.5cm 

\section{Introduction}

The fusion rule algebra \cite{BPZ,verl,gepwitt} have been studied extensively [2-10], 
since it seems to play an important role in the classification of the rational conformal field theories. In this paper we concentrate on the fusion rules of the extended chiral algebra, namely: we investigate the fusion rules of the WZW theory  \cite{gepwitt} on the non-simply connected manifolds $SU(n)/Z_{m}$ at levels\footnote{We will denote these levels generically by $k^{\star}$} 

$$k=n r \hskip 0.8cm {\rm for \hskip 0.3cm n \hskip 0.3cm odd}, \eqno(1.1.a)$$ 
$$k=2 n r \hskip 0.8cm {\rm for \hskip 0.3cm n \hskip 0.3cm even,}\eqno(1.1.b)$$since for these levels the chiral algebra is an extension of the usual current algebra. Here $Z_{m}$ is the subgroup of the center and $n=m p$, where p is a positive integer.

The primary fields of the WZW model may be labeled by the highest weight representations of the horizontal Lie algebra. Let the fundamental weights of $SU(n)$ be $\lambda_{i}$. Then the highest weights are given by $$\Lambda=\sum_{i=1}^{n-1}  m_{i} \lambda_{i},\eqno(1.2)$$where $m_{i}$ are nonnegative integers (Dynkin labels). The set of the integrable representations \cite{gepwitt} at the level $k$ is denoted by $P_{k}$,
 $$P_{k}=\{ m_{i}; 0 \le \sum_{i} m_{i} \le k \} \eqno(1.3)$$.

Let us denote by $\s{}{}$ the modular matrix describing the behavior of the characters of the ordinary diagonal $SU(n)_{k}$ theory under modular transformation $\tau \rightarrow -{1 \over \tau}$. The fusion coefficients of these models will be denoted by $N_{\Lambda_{1} \Lambda_{2}}^{\Lambda_{3}}$, where $\Lambda_{j} \in P_{k}$. The technic for the calculation of $N_{\Lambda_{1} \Lambda_{2}}^{\Lambda_{3}}$ was developed in \cite{gepwitt,gepri}. The modular matrix of the corresponding quotient $SU(n)/Z_{m}$ model will be denoted by $S$ without tilde and its fusion coefficients are denoted by $H_{\Lambda_{1} \Lambda_{2}}^{\Lambda_{3}}$. Let $\sigma$ be external automorphism and $C$ charge conjugation whose action on Dynkin labels is defined by:

$$\begin{array}{l} \sigma (m_1, m_2, ..., m_{n-1})=(k-\sum_{j} m_{j}, m_1,... , m_{n-2}), \\
C (m_1, m_2, ..., m_{n-1}) =(m_{n-1},m_{n-2},...,m_1)

\end{array} \eqno(1.4)$$The partition function of  $S(n)_{k^{\star}} /Z_{m}$ quotient models is given by

$$ Z=\sum_{{{\lambda \in P_{k^{\star}}} \atop {\lambda \hskip 0.1cm {\rm singlets \hskip 0.1cm of \hskip 0.1cm} Z_{m}}}} {n \over n_{\sigma \lambda}} |\sum_{\sigma} \chi_{\sigma \lambda}|^{2},\eqno(1.5)$$where $\sigma$ is external automorphism, $n_{\sigma \lambda}$ is the length of the orbit generated by the action of $\sigma$ on $\lambda$.

 Let $R$ be the fixed point of $\sigma$ {\it i.e.} $\sigma R=R$, so that at the levels $k=k^{\star}$ we have:

$$R={k^{\star} \over n} \rho, \eqno(1.6)$$where $\rho$ is the sum of fundamental weights of $SU(n)$

$$\rho=\sum_{i=1}^{n-1} \lambda_{i}={1 \over 2} \sum_{\alpha>0} \alpha,\eqno(1.7)$$where ${\alpha}$ are the simple roots of $SU(n)$.

For the reasons which will become clear later we will distinguish two cases: i) n-odd, ii) n-even.

\section{Fusion rules of $SU(n)_{n r} / Z_{n}$ for $n$ -odd}

As a first example we consider the case when $m=n$ and $n$ is odd prime number.

The generalized characters are given by $$ch_{\lambda}=\sum_{\sigma} \chi_{\sigma \lambda}, \hskip 0.8cm \lambda \not = R \eqno(2.1)$$the corresponding primary field is denoted by $\phi_{\lambda}$, where $\lambda$ is horizontal highest weight of some representative of the orbit generated by the action of $\sigma$ on $\lambda$. The character corresponding to the fixed point $R$ appears in the partition function with the multiplicity $n$, and in order to distinguish between $n$ corresponding primary fields we introduce additional quantum number $j=1,...,n$, these fields are denoted by $\phi_{R_{j}}$.

The modular matrix is given by \cite{zuber,yank1}

$$S={\pmatrix {n \s{\mu}{\nu}&\s{\mu }{\R{1}} & \s{\mu }{\R{2}} & ... & \s{\mu}{ \R{n}} \cr \s{\R{1}}{ \nu} &x&z&...&z \cr \s{\R{2}}{ \nu} &z&x&...&z \cr ...&...&...&...&... \cr \s{\R{n}}{ \nu} &z&z&...&x}},\eqno(2.2)$$where $\s{\mu }{\R{j}}=\s{\mu}{ R}$.  Note that the $S$ matrix given above is restricted to $Z_n$-singlets sector {\it i.e.} $\mu \lambda_{1}={\rm integer}.$

 It is easy to show that $\s{\mu}{R}=0$ when $\mu \lambda_{1} \not ={\rm integer}$, indeed using the relation \cite{gep1} $\s{\mu}{\sigma \nu}=e^{2 \pi i \mu \lambda_{1}} \s{\mu}{\nu}$ one has:

$$\s{\mu}{\sigma R} =e^{2 \pi i \mu \lambda_{1}} \s{\mu}{R}=\s{\mu}{R}=0, \hskip 0.8cm \mu \lambda_{1} \not ={\rm integer.} \eqno(2.3)$$ 

The previous observation enables us significantly to simplify the calculations: using the corresponding identities for the diagonal $SU(n)_{k^\star}$ models we project them onto $Z_{n}$ singlets sector, using Eq.(3). In order to calculate $x$ and $z$ we use the following equations:

$$(S T)^{3}=S^{2}=C, \hskip 0.5cm S S^{\dagger}=1, \eqno(2.4)$$where $C$ is charge conjugation matrix $(C^{2}=1)$ and $T$ is the matrix of the modular transformation $\tau \rightarrow \tau+1$ that is given by
$$T_{\mu,\nu}=\delta_{\mu,\nu} e^{2 \pi i (\Delta_{\mu}-{c \over 24})},\eqno(2.5)$$where $c$ is the central charge and $\Delta_{\mu}$ is the conformal weight of the primary field $\phi_{\mu}$ which is given by

$$\Delta_{\mu}={(\mu,\mu+2 \rho) \over 2 (k^{\star}+n)}. \eqno(2.6)$$

    More explicitly from Eqs.(4) we have:

$$\sum_{{\nu \not =R \atop \nu \lambda_{1}={\rm integer}} \atop {\nu \mod \sigma}} n \s{\mu}{\nu} \s{\nu}{R}=-(x+(n-1)z)\s{\mu}{R}=-\s{\mu}{R} \s{R}{R},\eqno(2.7)$$where the second equality is obtained from the similar equation for the diagonal $SU(n)_{k^{\star}}$ theory:

$$\sum_{\nu \not =R} \s{\mu}{\nu} \s{\nu}{R}+\s{\mu}{R} \s{R}{R}=0, \hskip 0.8cm \mu \not = R \eqno(2.8)$$and the fact that $\s{\mu}{R}=0$ when $\mu$ is not a singlet of $Z_{n}$\footnote{Note that the factor $n$ appears in the sum in the Eq.(2.7) when we switch the summation over $\lambda$ to be modulo $\sigma$}. Now from the Eq.(7) we have $$\s{R}{R}=x+(n-1) z. \eqno(2.9)$$ Taking the $R_{j}, R_{l}$ entry in the matrix relation $S^2=C$ and assuming that\footnote{This assumption is confirmed later, see Eq.(13) and below.} $[C]_{R_{j}, R_{l}}=\delta_{j,l}$ we have:

$$\sum_{{\nu \not =R} \atop {\nu \mod \sigma}} \s{R}{\nu} \s{\nu}{R}+x^{2}+(n-1) z^{2}=1, \eqno(2.10)$$
$$\sum_{{\nu \not =R} \atop {\nu \mod \sigma}} \s{R}{\nu} \s{\nu}{R}+2 x z +(n-2) z^2=0. \eqno(2.11)$$From the Eqs.(10-11) it follows that $$(x-z)^{2}=1. \eqno(2.12)$$Again taking the $R_{j},R_{l}$ component in the matrix relation $(S T)^{3}=C$ and subtracting from it the relation which should hold for the diagonal $SU(n)_{k^{\star}}$ theory, namely: 
$(\s{}{}{\tilde T})_{R,R}^3=1$ we have:

$$\s{R}{R}^3-n \sum_{w,q=1,...,n} S_{{R_{j}}{R_{w}}} S_{{R_{w}}{R_{q}}} S_{{R_{q}}{R_{l}}}=e^{i \pi ({c \over 4 }-6 \Delta_{R})}(1-n [C]_{R_{j}, R_{l}}), \eqno(2.13)$$where $\Delta_{R}$ is the conformal weight of the fields corresponding to the fixed point R and c is the central charge:

$$c={k (n^2-1) \over (k+n)}={r (n^2-1) \over (r+1)}, \eqno(2.14)$$ 

$$\Delta_{R}={(R,R+2 \rho)\over 2 (k+n) }={r (r+2) \rho^2 \over 2 (k+n) } ={r (r+2) (n^2-1) \over 24 (r+1)},\eqno(2.15)$$where in the last equality "strange" formula due to Freudenthal-de Vries was used\footnote{In the case of $SU(n)$ this formula looks like $\rho^2={ n (n^2-1) \over 12}$}. Using Eqs.(14-15) one may easily show that 

$$e^{i \pi ({c \over 4 }-6 \Delta_{R})}=e^{-i {\pi \over 4} r (n^2-1) }=1, \eqno(2.16)$$ for odd $n$. The left hand side of the Eq.(2.13) may be easily calculated:

$$\s{R}{R}^3-n \sum_{w,q=1,...,n} S_{{R_{j}}{R_{w}}} S_{{R_{w}}{R_{q}}} S_{{R_{q}}{R_{l}}}=\delta_{j,l} (x-z)^3+(1-\delta_{j,l})(n-1) (z-x)^3, \eqno(2.17)$$so that we conclude that $x-z=1$ and $[C]_{R_{j}, R_{l}}=\delta_{j,l}$ as was assumed before. Finally we will write down expressions for $x$ and $z$:
 
$$x={\s{R}{R}+n-1 \over n}, \hskip 0.8cm z={\s{R}{R}-1 \over n}. \eqno(2.18)$$

\vskip 0.8cm

 The part of fusion rules of the  $SU(n)_{n r} /Z_{n}$ quotient models is quite obvious:

$$\pr{\nu} \times \pr{\mu}=\sum_{{\eta \not = R \atop {\eta \mod \sigma}}} \sum_{\sigma} N_{\nu \mu}^{\sigma \eta} \pr{\eta} +N_{\nu \mu}^{R} \sum_{i=1}^{n} \pr{\R{i}}, \hskip 0.8cm \mu,\nu \not = R.\eqno(2.19)$$ 

The most interesting part are the fusion rules between the fields corresponding to the fixed point $R$. According to the Verlinde formula \cite{verl} the fusion coefficients are given by:

$$H_{R,R}^{\eta}=\sum_{{\mu \not =R} \atop {\mu \mod \sigma}} {{\s{R}{\mu}\s{R}{\mu}\s{\eta}{\mu}^{\dagger}} \over {\s{0}{\mu}} }+{{x^{2}+(n-1) z^{2}} \over {\s{0}{R}} }\s{\eta}{R}, \eqno(2.20)$$
$$H_{R,R^{\prime}}^{\eta}=\sum_{{\mu \not =R} \atop {\mu \mod \sigma}} {{\s{R}{\mu}\s{R}{\mu}\s{\eta}{\mu}^{\dagger}} \over {\s{0}{\mu}} }+{{2 x z+(n-2) z^{2}} \over {\s{0}{R}} }\s{\eta}{R}, \eqno(2.21)$$where $\eta \not =R$. From the Eqs.(12,20,21) one may find

$$H_{R,R}^{\eta}-H_{R,R^{\prime}}^{\eta}={{(x-z)^{2} \s{\eta}{R}^{\dagger}} \over \s{0}{R} }={{\s{\eta}{R}^{\dagger}} \over \s{0}{R} },\eqno(2.22)$$

$$H_{R,R}^{\eta}+(n-1) H_{R,R^{\prime}}^{\eta}=\sum_{{\mu \not =R} \atop {\mu \mod \sigma}} {{n \s{R}{\mu}\s{R}{\mu}\s{\eta}{\mu}^{\dagger}} \over {\s{0}{\mu}} }+{{(x+(n-1)z)^{2} \s{\eta}{R}^{\dagger}} \over \s{0}{R} }=$$

$$=\sum_{{\mu \not =R} \atop {\mu \mod \sigma}} {{n \s{R}{\mu}\s{R}{\mu}\s{\eta}{\mu}^{\dagger}} \over {\s{0}{\mu}} }+{{\s{R}{R}^{2} \s{\eta}{R}^{\dagger}} \over \s{0}{R}}=\sum_{\mu} {{n \s{R}{\mu}\s{R}{\mu}\s{\eta}{\mu}^{\dagger}} \over {\s{0}{\mu}} }=N_{R,R}^{\eta}, \hskip 0.8cm \eta \not = R. \eqno(2.23)$$Solving the Eqs. (2.22-2.23)  we find 

$$H_{R,R}^{\eta}={{N_{R,R}^{\eta}+f(\eta) (n-1) } \over n},\eqno(2.24)$$
$$H_{R,R^{\prime}}^{\eta}={{N_{R,R}^{\eta}-f(\eta)} \over n},\eqno(2.25)$$where we defined\footnote{This identity may be proved directly using the expression for the $S$ matrix and the denominator identity $\prod_{\alpha >0}(1-e^{-\alpha})^{mult \hskip 0.1cm \alpha}=\sum_{w \in W} \epsilon (w) e^{w(\rho)-\rho}$, where $W$ denotes Weyl group and $\epsilon (w)$ is the sign of Weyl reflection.} 

$$f(\eta)={{\s{\eta}{R}^{\dagger}} \over \s{0}{R}}=e^{{2 \pi i \over n}(\eta+2 \rho,\rho)} \prod_{\alpha >0} { 1-e^{{2 \pi i \over n}(\eta+\rho,\alpha)} \over 1-e^{{2 \pi i \over n}(\rho,\alpha)}}. \eqno(2.26)$$

The rest of the fusion coefficients may be calculated using the same trick:

$$H_{R,R}^{R}=\sum_{{\mu \not =R} \atop {\mu \mod \sigma}} {{\s{R}{\mu}\s{R}{\mu}\s{R}{\mu}^{\dagger}} \over {n \s{0}{\mu}} }+{{x^{3}+(n-1) z^{3}} \over {\s{0}{R}}},\eqno(2.27)$$

$$H_{R,R^{\prime}}^{R^{\prime \prime}}=\sum_{{\mu \not =R} \atop {\mu \mod \sigma}} {{\s{R}{\mu}\s{R}{\mu}\s{R}{\mu}^{\dagger}} \over {n \s{0}{\mu}} }+{{3 x z^{2}+(n-3) z^{3}}\over {\s{0}{R}}},\eqno(2.28)$$

$$H_{R,R}^{R^{\prime}}=\sum_{{\mu \not =R} \atop {\mu \mod \sigma}} {{\s{R}{\mu}\s{R}{\mu}\s{R}{\mu}^{\dagger}} \over {n \s{0}{\mu}} }+{{x^{2} z+ z^{2} x +(n-3) z^{3}}\over {\s{0}{R}}},\eqno(2.29)$$From Eqs.(27,28,29) we have

$$3 (n-1) H_{R,R}^{R^{\prime}}+(n-2) (n-1) H_{R,R^{\prime}}^{R^{\prime \prime}}+H_{R,R}^{R}=$$

$$=\sum_{{\mu \not =R} \atop {\mu \mod \sigma}} {{n \s{R}{\mu}\s{R}{\mu}\s{R}{\mu}^{\dagger}} \over {\s{0}{\mu}} }+{\s{R}{R}^{3} \over {\s{0}{R}}}=$$

$$=\sum_{\mu} {{\s{R}{\mu}\s{R}{\mu}\s{R}{\mu}^{\dagger}} \over {\s{0}{\mu}} }=N_{R,R}^{R}, \eqno(2.30)$$

$$H_{R,R}^{R}-H_{R,R}^{R^{\prime}}={x+z \over \s{0}{R}}, \hskip 0.8cm H_{R,R}^{R}-H_{R,R^{\prime}}^{R^{\prime \prime}}={x+2 z \over \s{0}{R}}. \eqno(2.31)$$Equations (30-31) determine $H_{R_{j},R_{l}}^{R_i}$ completely:

$$H_{R,R}^{R}={2-3 n +n^2-3 \s{R}{R}+3 n \s{R}{R} +N_{R,R}^{R} \s{0}{R} \over n^2 \s{0}{R}},\eqno(2.32)$$

$$H_{R,R}^{R^{\prime}}={2-n -3 \s{R}{R}+ n \s{R}{R} +N_{R,R}^{R} \s{0}{R} \over n^2 \s{0}{R}},\eqno(2.33)$$

$$H_{R,R^{\prime}}^{R^{\prime \prime} }={2-3 \s{R}{R}+N_{R,R}^{R} \s{0}{R} \over n^2 \s{0}{R}}.\eqno(2.34)$$

\vskip 3.0cm 

\subsection{Example: Fusion rules of $SU(3)_{k=3 r} /Z_{3}$}

In this case the condition $\lambda \lambda_{1}=$ integer, translates for the Dynkin labels into:

$$2 m_{1}+m_{2}=0 \mod 3. \eqno(2.35)$$The function $f(\eta) \equiv f(m_{1},m_{2})$ defined in Eq.(26) is given by:

$$f(m_{1},m_{2})=\left\{ \begin{array}{ll}
            1 & \mbox{for $m_{1}+m_{2}=3p$} \\
            0 & \mbox{for $m_{1}+m_{2}=3p+1$} \\
            -1 & \mbox{for $m_{1}+m_{2}=3p+2$}
 \end{array} \right., \eqno(2.36)$$where $p$ is a non-negative integer. Using the explicit expression for the fusion coefficients of the diagonal SU(3) models \cite{cum} one may show that at levels $k=3r$:

 $$N_{R,R}^{R}=N_{(r,r);(r,r)}^{(r,r)}=r+1. \eqno(2.37)$$By direct calculation one may show

$$\s{0}{R}={1 \over r+1}. \eqno(2.38)$$Substituting Eqs.(37,38) into Eq.(33) we find:

$$H_{R,R}^{R^{\prime}}=0, \eqno(2.39)$$

$$H_{R,R}^{R}={{N_{R,R}^{R}+2 f(R)  } \over 3},\eqno(2.40)$$
$$H_{R,R^{\prime}}^{R^{\prime \prime}}={{N_{R,R}^{R}-f(R)} \over 3}.\eqno(2.41)$$We checked by computer that this result indeed holds for $k=3,6,9,12,15$

$${\bf SU(3)_{3}/Z_{3}}$$ 

The partition function is given by:

$$Z=|\chi_{(0,0)}+\chi_{(0,3)}+\chi_{(3,0)}|^{2}+3 |\chi_{(1,1)}|^{2},\eqno(2.42)$$The correspondence between primary fields and generalized characters is shown below:

$$\begin{array}{lcc}
{\phi_{(0,0)}} & \leftrightarrow & {\chi_{(0,0)}+\chi_{(0,3)}+\chi_{(3,0)}} \\
{\phi_{(1,1)}}, {\phi_{(1,1)}^{\prime}}, {\phi_{(1,1)}^{\prime \prime}} & \leftrightarrow & \chi_{(1,1)} \end{array}\eqno(2.43)$$

Fusion rules are given by:

$$\begin{array}{l}
\phi_{(1,1)} \times  \phi_{(1,1)}  ={\bf 1}  \\
\phi_{(1,1)}^{\prime} \times  \phi_{(1,1)}^{\prime} ={\bf 1}  \\
\phi_{(1,1)}^{\prime \prime} \times  \phi_{(1,1)}^{\prime \prime} ={\bf 1}  \\
\phi_{(1,1)} \times  \phi_{(1,1)}^{\prime}=\phi_{(1,1)}^{\prime \prime}, etc.
\end{array}\eqno(2.44)$$

$${\bf SU(3)_{6}/Z_{3}}$$ 

$${\small Z=|\chi_{(0,0)}+\chi_{(0,6)}+\chi_{(6,0)}|^{2}+|\chi_{(1,1)}+\chi_{(1,4)}+\chi_{(4,1)}|^{2}+|\chi_{(3,3)}+\chi_{(3,0)}+\chi_{(0,3)}|^{2}+3 |\chi_{(2,2)}|^{2}}. \eqno(2.45)$$Primary fields:

\def\f{{\phi_{(1,1)}}}
\def\se{{\phi_{(3,3)}}}
\def\a{\phi_{(2,2)}}
\def\b{\phi^{\prime}_{(2,2)}}
\def\c{{\phi^{\prime \prime}_{(2,2)}}}

$$\begin{array}{lcc}
{\phi_{(0,0)}} & \leftrightarrow & {\chi_{(0,0)}+\chi_{(0,6)}+\chi_{(6,0)}} \\
{\phi_{(1,1)}} & \leftrightarrow & {\chi_{(1,1)}+\chi_{(1,4)}+\chi_{(4,1)}} \\
{\phi_{(3,3)}} & \leftrightarrow & {\chi_{(3,3)}+\chi_{(3,0)}+\chi_{(0,3)}} \\
{\phi_{(2,2)}}, {\phi_{(2,2)}^{\prime}}, {\phi_{(2,2)}^{\prime \prime}} & \leftrightarrow & \chi_{(2,2)} 
\end{array}\eqno(2.46)$$

The fusion rules are given by:

$$ \begin{array}{l}
    \f {\times} \f =1+2 {\f} +2 {\se}+\a+\b+\c \\                             
    \f {\times} \se =2 {\f}+{\se}+\a+\b+\c    \\
     \f {\times} \a =\f+\se+\b+\c             \\
      \f {\times} \b =\f+\se+\a+\c            \\
       \f {\times} \c =\f+\se+\a+\b            \\
      \se \times \se =1+\f+\se +\a+\b+\c    \\                                       \se \times \a =\f+\se+\a               \\
     \se \times \b =\f+\se+\b              \\                                        \se \times \c =\f+\se+\c               \\ 
      \a \times \a =1+\se+\a                \\
      \a \times \b =\f+\c, \hskip 0.35cm etc. 

\end{array}  \eqno(2.47)$$This fusion ring is equivalent to the fusion ring of the $SU(2)_{16}/Z_{2}$ see Sec.(4.1)

%%%%%%%%%%%%%%%%%%%%%%%%%%%%

$${\bf SU(3)_{9}/Z_{3}}$$

\def\po#1#2{\phi_{(#1,#2)}}
\def\lhs#1#2#3#4{\po{#1}{#2} \times \po{#3}{#4}}

The partition function is given by
{\small $$ Z=|\chi_{(0,0)}+\chi_{(0,9)}+\chi_{(9,0)}|^{2}+|\chi_{(1,1)}+\chi_{(1,7)}+\chi_{(7,1)}|^{2}+|\chi_{(2,2)}+\chi_{(2,5)}+\chi_{(5,2)}|^{2}+|\chi_{(0,3)}+\chi_{(6,0)}+\chi_{(3,6)}|^{2}+$$}

{\small $$ |\chi_{(3,0)}+\chi_{(0,6)}+\chi_{(6,3)}|^{2}+3 |\chi_{(3,3)}|^{2}. \eqno(2.48)$$}The primary fields are:

$$ \begin{array}{lcc}

{\phi_{(0,0)}} & \leftrightarrow & {\chi_{(0,0)}+\chi_{(0,9)}+\chi_{(9,0)}} \\
{\phi_{(1,1)}} & \leftrightarrow & {\chi_{(1,1)}+\chi_{(1,7)}+\chi_{(7,1)}} \\
{\phi_{(2,2)}} & \leftrightarrow & {\chi_{(2,2)}+\chi_{(2,5)}+\chi_{(5,2)}} \\
{\phi_{(0,3)}} & \leftrightarrow & {\chi_{(0,3)}+\chi_{(6,0)}+\chi_{(3,6)}} \\
{\phi_{(3,0)}} & \leftrightarrow & {\chi_{(3,0)}+\chi_{(0,6)}+\chi_{(6,3)}} \\
{\phi_{(3,3)}}, {\phi_{(3,3)}^{\prime}}, {\phi_{(3,3)}^{\prime \prime}} & \leftrightarrow & \chi_{(3,3)} 
\end{array} \eqno(2.49)$$Fusion rules are given by:

{\small $$ \begin{array}{l}
\lhs{1}{1}{1}{1}={\bf 1}+2\po{1}{1}+\po{2}{2}+\po{0}{3}+\po{3}{0} \\
\lhs{1}{1}{2}{2}=\po{1}{1}+2\po{2}{2}+\po{0}{3}+\po{3}{0}+\po{4}{4}+\po{3}{3}+\po{3}{3}^{\prime}+\po{3}{3}^{\prime \prime} \\
\lhs{1}{1}{0}{3}=\po{1}{1}+\po{2}{2}+\po{0}{3}+\po{4}{4} \\                                                         
\lhs{1}{1}{3}{0}=\po{1}{1}+\po{2}{2}+\po{3}{0}+\po{4}{4} \\                                                         
\lhs{1}{1}{4}{4}=2\po{2}{2}+\po{0}{3}+\po{3}{0}+2\po{4}{4}+\po{3}{3}+\po{3}{3}^{\prime}+\po{3}{3}^{\prime \prime} \\        \lhs{1}{1}{3}{3}= \po{2}{2}+\po{4}{4}+ \po{3}{3}^{\prime}+\po{3}{3}^{\prime \prime} \\                                      \lhs{2}{2}{2}{2}= {\bf 1}+2\po{1}{1}+5\po{2}{2}+2\po{0}{3}+2\po{3}{0}+5\po{4}{4}+2(\po{3}{3}+\po{3}{3}^{\prime}+\po{3}{3}^{\prime \prime}) \\                                              
\lhs{2}{2}{0}{3}=\po{1}{1}+2\po{2}{2}+\po{0}{3}+\po{3}{0}+2\po{4}{4}+\po{3}{3}+\po{3}{3}^{\prime}+\po{3}{3}^{\prime \prime} \\
\lhs{2}{2}{3}{0}=\po{1}{1}+2\po{2}{2}+\po{0}{3}+\po{3}{0}+2\po{4}{4}+\po{3}{3}+\po{3}{3}^{\prime}+\po{3}{3}^{\prime \prime} \\                                               
\lhs{2}{2}{4}{4}=2\po{1}{1}+5\po{2}{2}+2(\po{0}{3}+\po{3}{0})+4\po{4}{4}+2(\po{3}{3}+\po{3}{3}^{\prime}+\po{3}{3}^{\prime \prime}) \\                                                
\lhs{2}{2}{3}{3}=\po{1}{1}+2\po{2}{2}+\po{0}{3}+\po{3}{0}+2\po{4}{4}+\po{3}{3}+\po{3}{3}^{\prime}+\po{3}{3}^{\prime \prime} \\                                                 
                                                
\lhs{0}{3}{0}{3}= \po{2}{2}+2 \po{3}{0}+ \po{4}{4} \\                                             
\lhs{0}{3}{3}{0}= {\bf 1}+ \po{1}{1}+\po{2}{2}+\po{3}{3}+\po{3}{3}^{\prime}+\po{3}{3}^{\prime \prime} \\                                               
\lhs{0}{3}{4}{4}=\po{1}{1}+2\po{2}{2}+\po{3}{0}+2\po{4}{4}+\po{3}{3}+\po{3}{3}^{\prime}+\po{3}{3}^{\prime \prime} \\
                                               
\lhs{0}{3}{3}{3}=\po{2}{2}+\po{0}{3}+\po{4}{4}+\po{3}{3} \\
                                               
\lhs{3}{0}{3}{0}=\po{2}{2}+2 \po{0}{3}+ \po{4}{4} \\                                                 
\lhs{3}{0}{4}{4}=\po{1}{1}+2\po{2}{2}+\po{0}{3}+2\po{4}{4}+\po{3}{3}+\po{3}{3}^{\prime}+\po{3}{3}^{\prime \prime}  \\                                              
\lhs{3}{0}{3}{3}=\po{2}{2}+\po{3}{0}+\po{4}{4}+\po{3}{3} \\
                                                
\lhs{4}{4}{4}{4}= {\bf 1}+2\po{1}{1}+4\po{2}{2}+2(\po{0}{3}+\po{3}{0})+4\po{4}{4}+\po{3}{3}+\po{3}{3}^{\prime}+\po{3}{3}^{\prime \prime} \\
                                               
\lhs{4}{4}{3}{3}=\po{1}{1}+2\po{2}{2}+\po{0}{3}+\po{3}{0}+2\po{4}{4}+\po{3}{3}^{\prime}+\po{3}{3}^{\prime \prime} \\
                                                
\lhs{3}{3}{3}{3}={\bf 1}+ \po{2}{2}+\po{0}{3}+\po{3}{0}+2\po{3}{3} \\                                               

\po{3}{3} \times \po{3}{3}^{\prime}=\po{1}{1}+\po{2}{2}+\po{4}{4}+\po{3}{3}^{\prime \prime}

\end{array}  \eqno(2.50)$$}The fusion rules of $SU(3)_{12}/Z_3$ models are given in the appendix.

\vskip 0.5cm

The graphs whose incidence matrix is equal to $N_{\f}$ are shown at the Fig.[1], where $\f$ corresponds to the character $\chi_{(1,1)}+\chi_{(k-2,1)}+\chi_{(1,k-2)}$.

\newpage

$$\pscaption{\psboxto(5cm;4cm){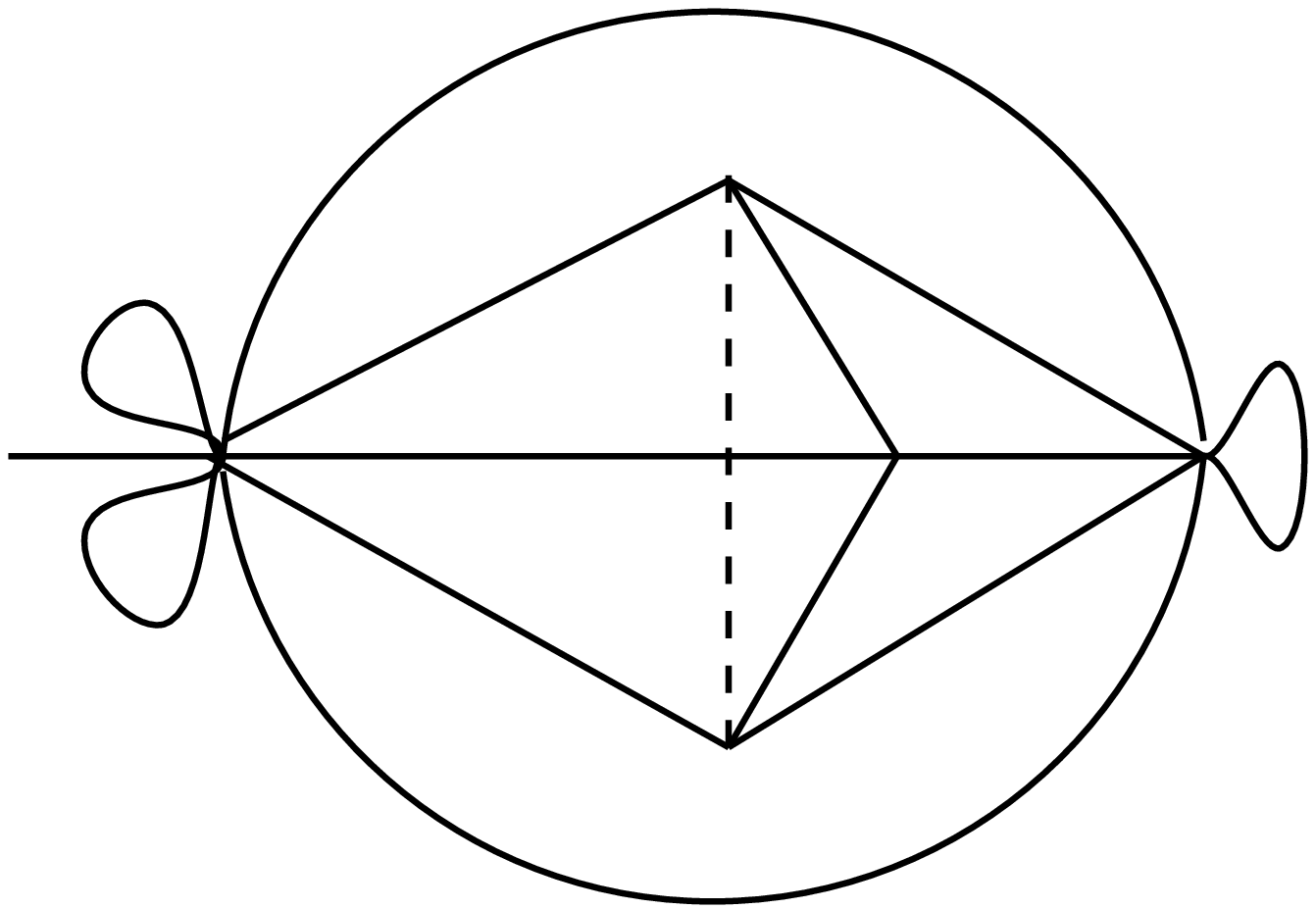}}{$k=6$} \hskip 0.5cm \pscaption{\psboxto(5cm;4cm){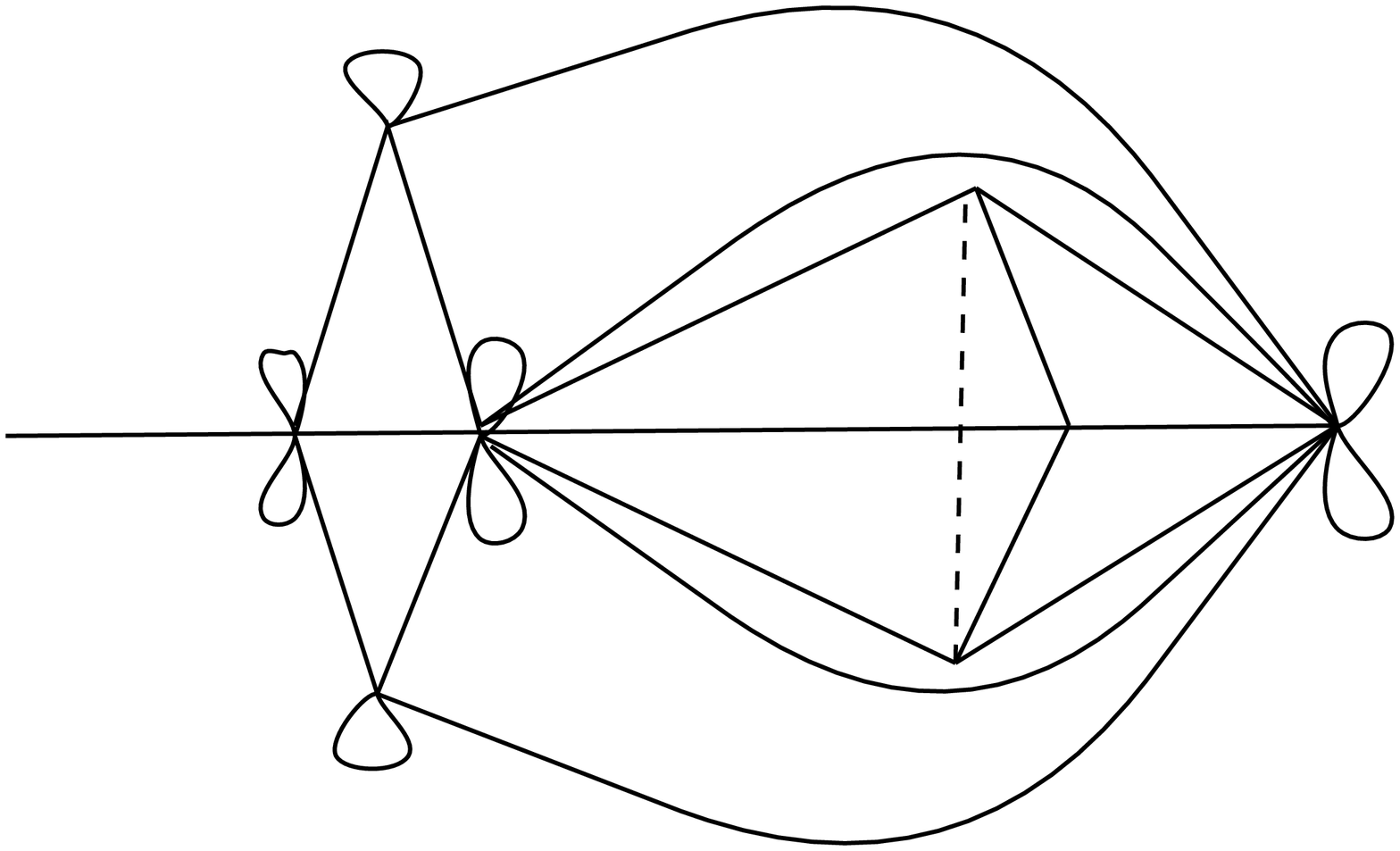}}{$k=9$}$$

$$\pscaption{\psboxto(6cm;5cm){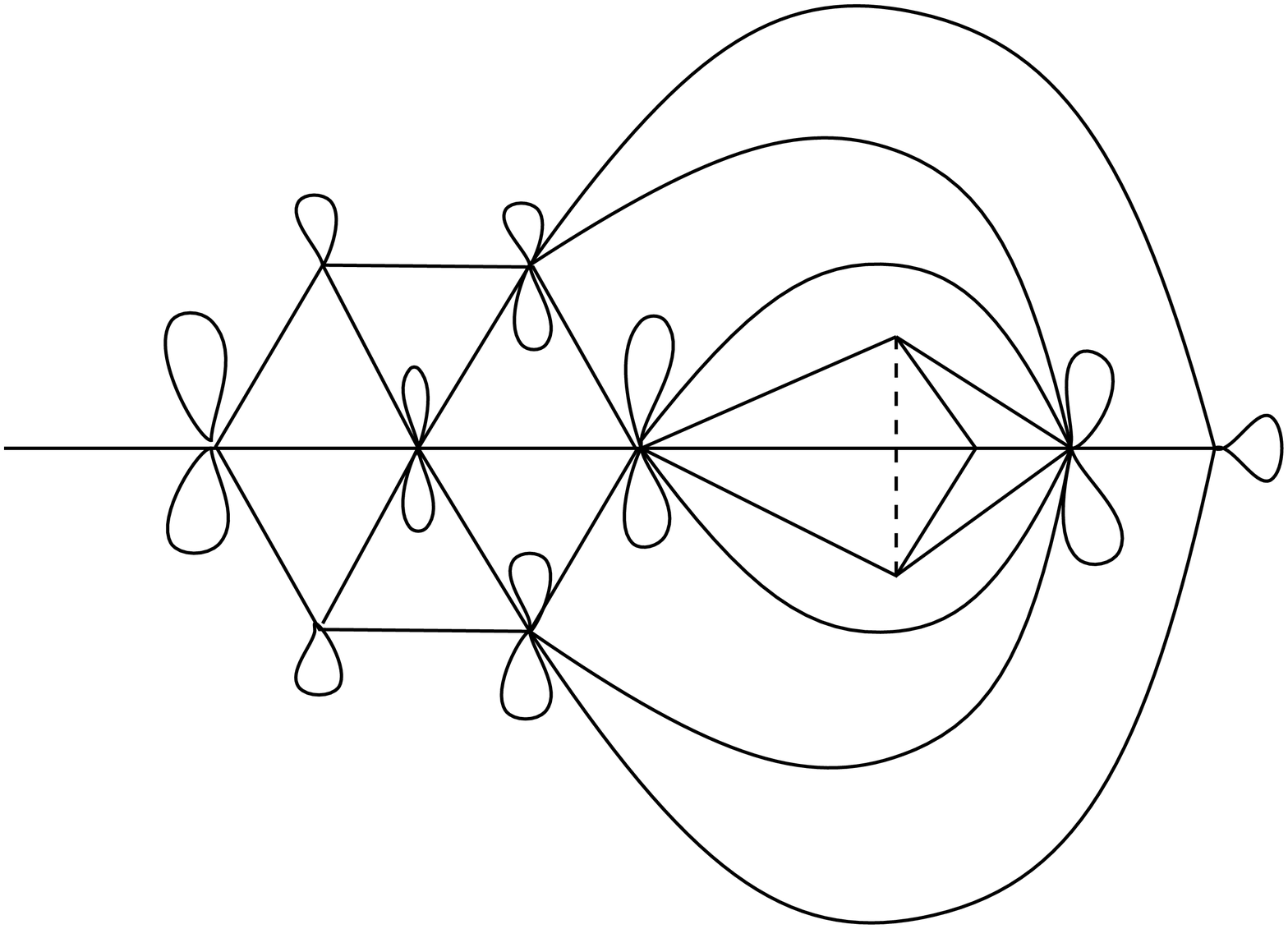}}{$k=12$} \hskip 0.5cm \pscaption{\psboxto(6cm;5.5cm){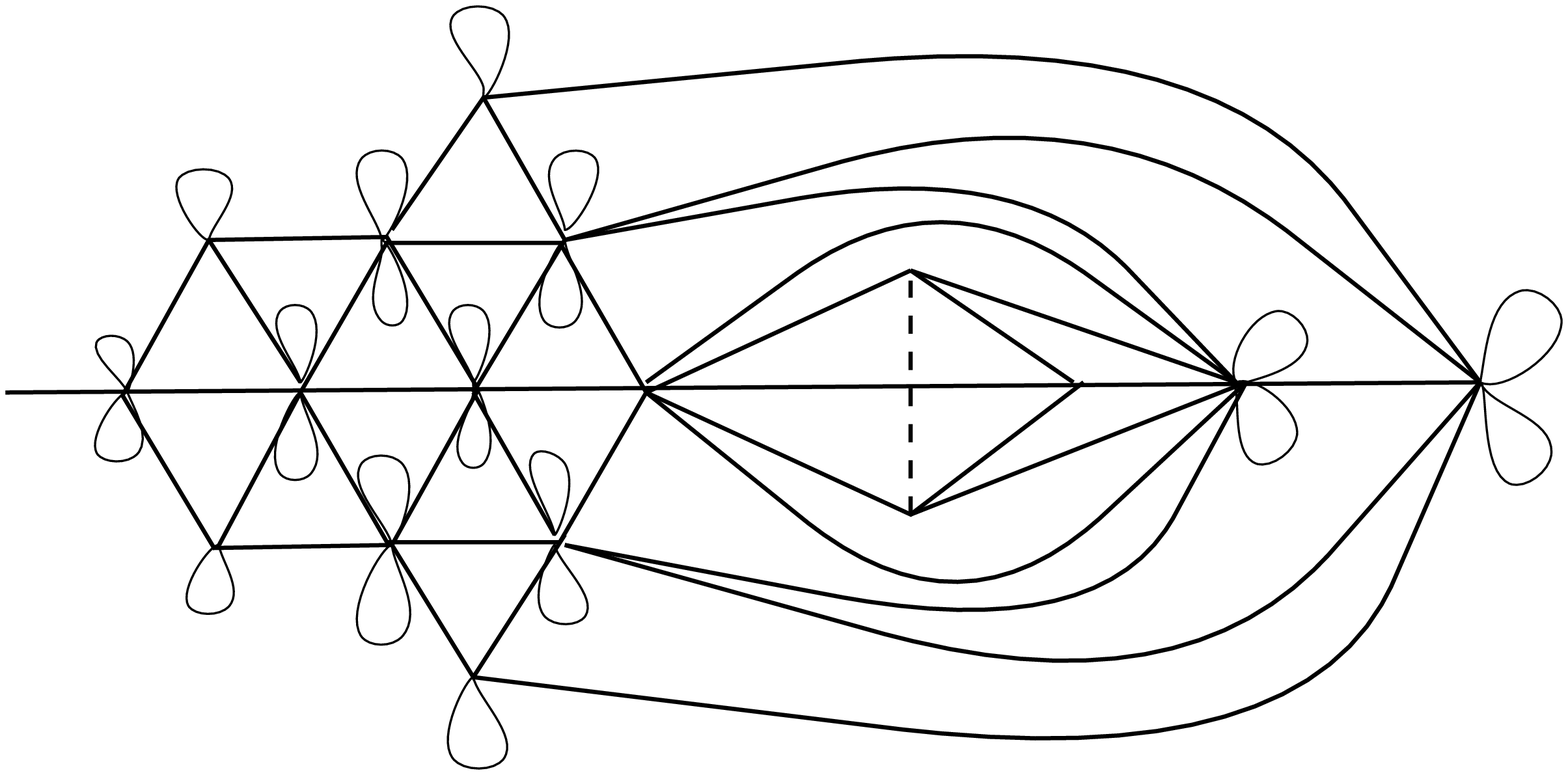}}{$k=15$}$$

Fig.1: Fusion graphs of the $SU(3)_{k=3 r}/Z_{3}$ models. Here the left free end of the graphs corresponds to the identity operator ${\bf 1}$, charge conjugation $C(m_{1},m_{2})=(m_{2},m_{1})$ corresponds to the 
reflection under the horizontal line. The fields $\phi_{R_{j}}$, $j=1,2,3$ are sitting on the common base of the two pyramids inside the internal circle.

\newpage

\section{Generalization for the arbitrary odd m}

\def\sitld{\tilde \sigma}

If $n$ is not a prime number there is an additional complication which may arise due to appearance of the representations $\zeta$  with 

$$1<n_{\sigma \zeta}<n, \eqno(3.1)$$it means that corresponding character will appear in the partition sum with the multiplicity ${n \over n_{\sigma \zeta}}$ and there will be ${n \over n_{\sigma \zeta}}$ primary fields corresponding to the same generalized character $\sum_{\sitld \in Z_{m}} \chi_{\sitld \zeta}$. In this case the matrix $S$ will be given by:

$$S=\pmatrix{m \s{\mu}{\nu} &n_{\sigma \zeta} \s{\mu}{\zeta_{b}} & \s{\mu}{R_{j}} \cr n_{\sigma \zeta}  \s{\zeta_{a}}{\nu} & y_{a,b} & {n_{\sigma \zeta} \over n} \s{\zeta}{R} \cr \s{R_{i}}{\nu}& {n_{\sigma \zeta} \over n} \s{R}{\zeta}& x_{i,j}}, \eqno(3.2)$$where $a,b=1,..,{n \over n_{\sigma \zeta}}$ and $i,j=1,...,n$. The fusion rules between primary fields $\phi_{\zeta_{a}}$ may be calculated in a way similar to what was done for the primary fields corresponding to the fixed point $R$. Note that the  equations for $x$, $z$ are the same so that fusion rules between fields $\phi_{R_{i}}$ are remained unchanged.

\section{Fusion rules of $SU(n)_{2 n r} / Z_{m}$ for $n$ -even}

For the even $n$ at the level $k=2 n r$ Eq.(2.16) leads us to

$$e^{i \pi ({c \over 4 }-6 \Delta_{R})}=e^{-i {\pi \over 2} r }. \eqno(4.1)$$For $r$ even which is divisible by 4 the right hand side of the Eq.(4.1) is equal to 1, and the result is identical to the discussed in the case of odd $n$. For  $r$ even which is not divisible by 4 from Eq.(2.17) one may find $x-z=-1$, and the only difference from the previously discussed result is in the expressions for the $x$ and $z$ which lead to the different expression for the fusion coefficients $H_{R_{j},R_{l}}^{R_{w}}$,

$$\begin{array}{l}
 H_{R,R}^{R}={-2+3 m -m^2-3 \s{R}{R}+3 m \s{R}{R} +N_{R,R}^{R} \s{0}{R} \over m^2 \s{0}{R}}   \\
H_{R,R}^{R^{\prime}}={-2+m -3 \s{R}{R}+ m \s{R}{R} +N_{R,R}^{R} \s{0}{R} \over m^2 \s{0}{R}}   \\
H_{R,R^{\prime}}^{R^{\prime \prime} }={-2-3 \s{R}{R}+N_{R,R}^{R} \s{0}{R} \over m^2 \s{0}{R}}  
        
\end{array}. \eqno(4.2)$$

\vskip 0.7cm

If $r$ is odd one may show that Eqs.(2.4) do not have solutions for $m>2$. Indeed denoting:

$$S_{R_{j},R_{l}}=x_{j,l}, \eqno(4.3)$$where we anticipate that the naive parameterization of the modular matrix $S$ which we used in Eq.(2.2) (only $x$ and $z$) will not work.  Using $S^2=C$ we have: 

$$\sum_{\zeta=1}^{m} x_{j,\zeta} x_{\zeta,l}={\s{R}{R}^{2}-1 \over m}+C_{\R{j}, \R{l}},\eqno(4.4)$$plugging this into Eq.(2.4) we obtain:

$$\s{R}{R}-m \sum_{\eta} C_{\R{j}, \R{\eta}} x_{\eta,l}=i (-1)^{{r+1} \over 2} (1-m C_{\R{j}, \R{\l}} ). \eqno(4.5)$$Calculating $x_{\eta,l}$ from this equation we find:
$$x_{\eta,l}={{\s{R}{R}-i (-1)^{{r+1} \over 2} (1-m \delta_{\eta,l})} \over m}. \eqno(4.6)$$Using Eq.(4) and Eq.(6) we get the following equality:

$$C_{\R{\zeta}, \R{\eta}}={2 \over m}-\delta_{\zeta,\eta}, \eqno(4.7)$$which shows that the only case which may be treated as reasonable solution is when $m=2$, since when $m \not = 2$ the fusion coefficients $N_{i,j}^{0}$ are fractional.

 In order to illustrate what happens
 we will consider the simplest example corresponding to $SU(2)_{4 r} /Z_{2}$, the generalization for arbitrary even $n$ and $m=2$ is obvious from the previous discussion. 

\subsection{Example: $SO(3)_{4 r}=SU(2)_{4 r} /Z_{2}$ } 
$${\psboxto(7cm;3cm){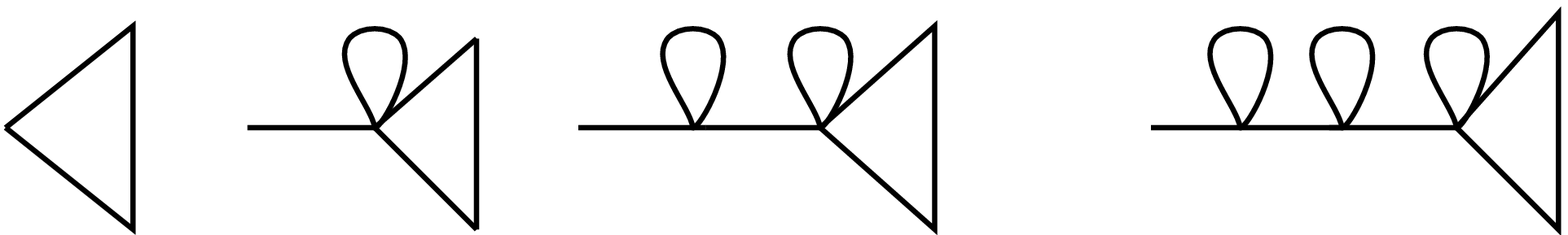}}$$

 {{\small Fig.2: Fusion graphs for $SU(2)_{4 r} /Z_{2}$, whose incidence matrix is given by $N_{\phi_{1}}$}}
\vskip 0.8cm

In this case the matrix $S$ is given by:

$$S={\pmatrix{2 \s{\mu}{\nu}&\s{\mu}{\R{1}} &\s{\mu}{\R{2}} \cr \s{\R{1}}{\nu} &x&z \cr \s{\R{2}}{\nu} &z&x}},\eqno(4.8)$$where $x$ and $z$ are given by Eq.(6). Labeling primary fields by their isospin $j={m_{1} \over 2}$ and repeating the same steps as in the Sec.(2) we arrive to the following expression for the function $f(m_{1})$:

$$f(j)=e^{i \pi j}. \eqno(4.9)$$

The fusion rules are given by:

\def\f#1{\phi_{#1}}
\def\m{\times}

$$\begin{array}{l}

\f{i} \m \f{j}=\sum_{p \not = r} (N_{i,j}^{p}+N_{i,j}^{2 r-p}) \f{p}+N_{i,j}^{r} (\f{r}+\f{r}^{\prime}) \\
             \\

\f{r} \m \f{r}=\sum_{p=0}^{r \over 2} \f{2 p} \hskip 1cm {\rm r-even} \\

\f{r} \m \f{r}^{\prime}=\sum_{p=0}^{{r-2} \over 2} \f{2 p+1} \hskip 1cm {\rm r-even} \\

\f{r} \m \f{r}=\sum_{p=0}^{{{r-3} \over 2}} \f{2 p+1}+\f{r}^{\prime} \hskip 1cm {\rm r-odd} \\
                                                                                           \\

\f{r} \m \f{r}^{\prime}=\sum_{p=0}^{{r-1} \over 2} \f{2 p} \hskip 1cm {\rm r-odd} 

\end{array}. \eqno(4.10)$$For $r=4$ this result was obtained in \cite{verl1} and it is in agreement with our result up to a type error. Note that the fusion ring when $r=4$ is identical to the fusion ring of $SU(3)_{6} / Z_{3}$.

\vskip 0.6cm
\hskip 0.8cm {\bf Note added:}
While writing this work we received \cite{fuchs1} which overlaps part of the discussion here.

\vskip 0.6cm

\hskip 0.8cm {\bf Acknowledgments}
\vskip 0.3cm

We thank S. Cherkis and A. Kapustin for useful discussions.

\newpage

\hskip 0.8cm {\bf APPENDIX}

%%%%%%%%%%%%%%%%%%%%%%%%%%%%%%%%%%%%%%%%%%%%%%%%%%%%%%%%%%%%%%%%%%%%%%%

$${\bf SU(3)_{12}/Z_{3}}$$ 

\def\q{\phi_{1}}
\def\w{\phi_{2}}
\def\e{\phi_{3}}
\def\r{\phi_{4}}
\def\t{\phi_{5}}
\def\y{\phi_{6}}
\def\u{\phi_{7}}
\def\i{\phi_{8}}
\def\o{\phi_{9}}
\def\p{\phi_{10}}
\def\pp{\phi_{10}^{\prime}}
\def\ppp{\phi_{10}^{\prime \prime}}

The partition function is given by:

 $$\begin{array}{c}

Z=|\chi_{(0,0)}+\chi_{(0,12)}+\chi_{(12,0)}|^{2}+|\chi_{(1,1)}+\chi_{(1,10)}+\chi_{(10,1)}|^{2}+|\chi_{(2,2)}+\chi_{(2,8)}+\chi_{(8,2)}|^{2}+\\

|\chi_{(3,3)}+\chi_{(3,6)}+\chi_{(6,3)}|^{2}+|\chi_{(0,3)}+\chi_{(3,9)}+\chi_{(9,0)}|^{2}+|\chi_{(3,0)}+\chi_{(9,3)}+\chi_{(0,9)}|^{2}+\\

|\chi_{(1,4)}+\chi_{(4,7)}+\chi_{(7,1)}|^{2}+|\chi_{(4,1)}+\chi_{(7,4)}+\chi_{(1,7)}|^{2}+|\chi_{(5,5)}+\chi_{(2,5)}+\chi_{(5,2)}|^{2}+\\

 |\chi_{(6,6)}+\chi_{(0,6)}+\chi_{(6,0)}|^{2}+3 |\chi_{(4,4)}|^{2}. 

\end{array} \eqno(..)$$The primary fields are:
$$ \begin{array}{lcc}

{\bf 1} & \leftrightarrow & {\chi_{(0,0)}+\chi_{(0,12)}+\chi_{(12,0)}} \\
{\q} & \leftrightarrow & {\chi_{(1,1)}+\chi_{(1,10)}+\chi_{(10,1)}} \\
{\w} & \leftrightarrow & {\chi_{(2,2)}+\chi_{(8,2)}+\chi_{(2,8)}} \\
{\e} & \leftrightarrow & {\chi_{(3,3)}+\chi_{(3,6)}+\chi_{(6,3)}} \\
{\r}  & \leftrightarrow & {\chi_{(0,3)}+\chi_{(3,9)}+\chi_{(9,0)}} \\
{\t}  & \leftrightarrow & {\chi_{(3,0)}+\chi_{(9,3)}+\chi_{(0,9)}} \\
{\y}  & \leftrightarrow & {\chi_{(1,4)}+\chi_{(4,7)}+\chi_{(7,1)}} \\
{\u}   & \leftrightarrow & {\chi_{(4,1)}+\chi_{(7,4)}+\chi_{(1,7)}} \\
 {\i}  & \leftrightarrow & {\chi_{(5,5)}+\chi_{(2,5)}+\chi_{(5,2)}} \\
  {\o} & \leftrightarrow & {\chi_{(6,6)}+\chi_{(0,6)}+\chi_{(6,0)}} \\
 {\p},{\pp},{\ppp}    & \leftrightarrow & {\chi_{(4,4)}}
\end{array}$$

{\small $$ \begin{array}{l}

\q \times \q={\bf 1}+2\q+\w+\r+\t \\
\q \times \w=\q+2\w+\e+\r+\t+\y+\u \\
\q \times \e=\w+2\e+\y+\u+2\i+\p+\pp+\ppp \\
\q \times \r=\q+\w+\r+\y \\
\q \times \t=\q+\w+\t+\u \\
\q \times \y=\w+\e+\r+2\y+\i+\o \\
\q \times \u=\w+\e+\t+2\u+\i+\o \\
\q \times \i=2\e+\y+\u+2\i+\o+\p+\pp+\ppp \\
\q \times \o=\y+\u+\i+\o \\
\q \times \p=\e+\i+\pp+\ppp \\
\w \times \w={\bf 1}+2\q+3\w+2\e+\r+\t+2(\y+\u)+2\i+2\o+\p+\pp+\ppp \\
\w \times \e=\q+2\w+5\e+\r+\t+3(\y+\u)+2\i+2\o+2(\p+\pp+\ppp) \\
\w \times \r=\q+\w+\e+\r+\t+\y+\u+\i \\
\w \times \t=\q+\w+\e+\r+\t+\y+\u+\i \\

\w \times \y=\q+2\w+3\e+\r+\t+2(\y+\u)+3\i+\o+\p+\pp+\ppp \\
\w \times \u=\q+2\w+3\e+\r+\t+2(\y+\u)+3\i+\o+\p+\pp+\ppp \\
\w \times \i=2\w+5\e+\r+\t+3(\y+\u)+5\i+\o+2(\p+\pp+\ppp) \\
\w \times \o=2\w+2\e+\y+\u+\i+\o+\p+\pp+\ppp \\
\w \times \p=\po{2}{2}+2\e+\y+\u+2\i+\o+\p+\pp+\ppp \\

\e \times \e={\bf 1}+2\q+5\w+8\e+2(\r+\t)+5(\y+\u)+8\i+3\o+3(\p+\pp+\ppp )\\
\e \times \r=\w+2\e+\r+\y+\u+2\i+\o+\p+\pp+\ppp \\
\e \times \t=\w+2\e+\t+\y+\u+2\i+\o+\p+\pp+\ppp \\
\e \times \y=\q+3\w+5\e+\r+\t+4\y+3\u+5\i+2\o+2(\p+\pp+\ppp) \\
\e \times \u=\q+3\w+5\e+\r+\t+3\y+4\u+5\i+2\o+2(\p+\pp+\ppp) \\
\e \times \i=2\q+5\w+8\e+2(\r+\t)+5(\y+\u)+7\i+3\o+3(\p+\pp+\ppp) \\
\e \times \o=2\w+3\e+\r+\t+2(\y+\u)+3\i+\o+\p+\pp+\ppp \\
\e \times \p=\q+2\w+3\e+\r+\t+2(\y+\u)+3\i+\o+2\p+\pp+\ppp \\
\r \times \r=\w+\t+\y+\o \\
\r \times \t={\bf 1}+\q+\w+\e \\
\r \times \y=\w+\e+\y+2\u+\i+\o \\
\r \times \u=\q+\w+\e+\t+\u+\i+\p+\pp+\ppp \\
\r \times \i=\w+\e+\y+\u+2\i+\o+\p+\pp+\ppp \\
\r \times \o=\e+\t+\u+\i \\
\r \times \p =\e+\y+\i+\p \\
\t \times \t =\w+\r+\u+\o \\
\t \times \y = \q+\w+\e+\r+\y+\i+\p+\pp+\ppp \\
\t \times \u =\w+\e+2\y+\u+\i+\o  \\
\t \times  \i=\w+2\e+\y+\u+2\i+\p+\pp+\ppp \\
\t \times  \o=\e+\r+\y+\i   \\
\t \times  \p =\e+\u+\i+\p  \\
\y \times  \y=2\w+3\e+2\t+\y+4\u+3\i+2\o+\p+\pp+\ppp  \\
\y \times  \u={\bf 1}+2\q+2\w+4\e+\r+\t+\y+\u+3\i+2(\p+\pp+\ppp)  \\
\y \times   \i=\q+3\w+5\e+\r+\t+3(\y+\u)+5\i+2\o+2(\p+\pp+\ppp) \\
\y \times  \o=\q+\w+2\e+\t+2\u+2\i+\p+\pp+\ppp  
  
\end{array} $$}

{\small $$ \begin{array}{l} 
\y \times  \p=\w+2\e+\r+2\y+\u+2\i+\o+\pp+\ppp \\
\u \times \u=2\w+3\e+2\r+4\y+\u+3\i+2\o+\p+\pp+\ppp \\
\u \times \i=\q+3\w+5\e+\r+\t+3(\y+\u)+5\i+2\o+2(\p+\pp+\ppp) \\
\u \times \o=\q+\w+2\e+\r+2\y+2\i+\p+\pp+\ppp  \\
\u \times \p=\w+2\e+\t+\y+2\u+2\i+\o+\pp+\ppp \\
\i \times \i ={\bf 1}+2\q+5\w+7\e+2(\r+\t)+5(\y+\u)+7\i+3\o+3(\p+\pp+\ppp) \\
\i \times \o=\q+\w+3\e+\r+\t+2(\y+\u)+3\i+\o+\p+\pp+\ppp \\
\i \times \p=\q+2\w+3\e+\r+\t+2(\y+\u)+3\i+\o+\p+\pp+\ppp \\
\o \times \o={\bf 1}+\q+\w+\e+\i+\o+\p+\pp+\ppp  \\
\o \times \p=\w+\e+\y+\u+\i+\o+\p \\
\p \times \p={\bf 1}+\w+2\e+\r+\t+\i+\o+\p \\
\p \times \pp=\q+\w+\e+\y+\u+\i+2\ppp

\end{array}$$}

\end{document}